\documentclass{aa}
\usepackage{graphicx}
\usepackage{natbib}
\usepackage{psfig}
\newcommand\eg{{\it e.g.} }
\newcommand\ie{{\it i.e.} }
\newcommand\etal{et~al.}
\newcommand\HI{\ion{H}{I}}
\newcommand\HII{\ion{H}{II}}
\newcommand\Lya{Ly$\alpha$}
\newcommand\NV{\ion{N}{V}~$\lambda$~1240}
\newcommand\CIV{\ion{C}{IV}~$\lambda$~1549}
\newcommand\HeII{\ion{He}{II}~$\lambda$~1640}
\newcommand\CIII{\ion{C}{III]}~$\lambda$~1909}
\newcommand\OII{\ion{[O}{II]}~$\lambda$~3727}
\newcommand\OIII{\ion{[O}{III]}~$\lambda$~5007}
\newcommand\Hbeta{H$\beta$}
\newcommand\kms{\ifmmode {\rm\,km\,s^{-1}}\else${\rm\,km\,s^{-1}}$\fi}
\font\aipsfont = cmsy9 scaled\magstep1

\newcommand\aips {{\aipsfont AIPS}}
\def\spose#1{\hbox to 0pt{#1\hss}}
\newcommand\simlt{\mathrel{\spose{\lower 3pt\hbox{$\mathchar"218$}}
     \raise 2.0pt\hbox{$\mathchar"13C$}}}
\newcommand\simgt{\mathrel{\spose{\lower 3pt\hbox{$\mathchar"218$}}
     \raise 2.0pt\hbox{$\mathchar"13E$}}}
\begin{document}
\title{CO emission and associated \HI\ absorption from a massive gas reservoir surrounding the $z=3$ radio galaxy B3~J2330+3927\thanks{Based on observations obtained with the IRAM Plateau de Bure Interferometer, the W.\ M.\ Keck telescope, the Westerbork Synthesis Radio Telescope and the Very Large Array.}}
\titlerunning{CO emission and \HI\ absorption in B3~J2330+3927.}

\author{Carlos De Breuck\inst{1}\thanks{Marie Curie fellow} \and Roberto Neri\inst{2} \and Raffaella Morganti\inst{3} \and Alain Omont\inst{1} \and Brigitte Rocca-Volmerange\inst{1} \and Daniel Stern\inst{4} \and Michiel Reuland\inst{5} \and Wil van Breugel\inst{5} \and Huub R\"ottgering\inst{6} \and S.\ A.\ Stanford\inst{5,7} \and Hyron Spinrad\inst{8} \and Mario Vigotti\inst{9} \and Melvyn Wright\inst{10}}

\offprints{Carlos De Breuck}
\institute{Institut d'Astrophysique de Paris, CNRS, 98bis Boulevard Arago, F-75014 Paris, France\\ \email{debreuck,omont,rocca@iap.fr} \and Institut de Radioastronomie Millim\'etrique, 300 rue de la piscine, F-38406 St. Martin-d'H\`eres, France \\  \email{neri@iram.fr} \and ASTRON, Postbus 2, NL-7990 AA Dwingeloo, The Netherlands \\ \email{morganti@nfra.nl} \and Jet Propulsion Laboratory, California Institute of Technology, Mail Stop 169-327, Pasadena, CA 91109, USA \\ \email{stern@zwolfkinder.jpl.nasa.gov} \and IGPP/LLNL, L-413, 7000 East Ave, Livermore, CA 94550, USA\\ \email{mreuland,wil@igpp.ucllnl.org} \and Sterrewacht Leiden, Postbus 9513, NL-2300 RA Leiden, The Netherlands\\ \email{rottgeri@strw.leidenuniv.nl} \and Department of Physics, University of California, Davis, CA 95616, USA\\ \email{adam@igpp.ucllnl.org}  \and Department of Astronomy, University of California, Berkeley, CA 94720, USA\\ \email{spinrad@astro.berkeley.edu} \and Istituto di Radioastronomia del CNR, Via Gobetti 101, I-40129 Bologna, Italy\\ \email{vigotti@ira.bo.cnr.it} \and Radio Astronomy Laboratory, University of California, Berkeley, CA 94720, USA\\ \email{mwright@astro.berkeley.edu}}

\date{Received 2002 August 12; accepted 2003 February 7}

\abstract{
We present results of a comprehensive multi-frequency study of the radio galaxy B3~J2330+3927.
The 1\farcs9 wide radio source, consisting of 3 components, is bracketed by 2 objects in our Keck $K-$band image. Optical and near-IR Keck spectroscopy of these two objects yield $z=3.087\pm 0.004$. The brightest ($K=18.8$) object has a standard type~II AGN spectrum, and is the most likely location of the AGN, which implies a one-sided jet radio morphology.\\
Deep 113~GHz observations with the IRAM Plateau de Bure Interferometer reveal CO~$J=4-3$ emission, which peaks at the position of the AGN.
The CO line is offset by $500$~\kms\ from the systemic redshift of the AGN, but corresponds very closely to the velocity shift of an associated \HI\ absorber seen in \Lya. This strongly suggests that both originate from the same gas reservoir surrounding the AGN host galaxy.
Simultaneous 230~GHz interferometer observations find a $\sim 3 \times$ lower integrated flux density when compared to single dish 250~GHz observations with MAMBO at the IRAM 30m telescope. This can be interpreted as spatially resolved thermal dust emission at scales of 0\farcs5 to 6\arcsec. \\
Finally, we present a $\tau < 1.3$\% limit to the \HI\ 21~cm absorption against the radio source, which represents the seventh non-detection out of 8 $z>2$ radio galaxies observed to date with the WSRT.\\
We present mass estimates for the atomic, neutral, and ionized hydrogen, and for the dust, ranging from $M(\HI)=2\times 10^7$~M$_{\odot}$ derived from the associated \HI\ absorber in \Lya\ up to $M({\rm H}_2)=7\times 10^{10}$~M$_{\odot}$ derived from the CO emission. This indicates that the host galaxy is surrounded by a massive reservoir of gas and dust. The $K-$band companion objects may be concentrations within this reservoir, which will eventually merge with the central galaxy hosting the AGN.\\
}

\maketitle
%

\section{Introduction}
The study of high redshift galaxies has for a long time been limited to the optical/near-IR and radio domains. Recent technological progress has now opened up other wavelength regimes (sub-mm, X-ray and even $\gamma$-ray) to high redshift studies.
Together with the discovery of optically-selected 'Lyman break' galaxies \citep[\eg][]{ste96,ste99,ouc01}, this has much broadened our knowledge of high redshift objects, which was previously based on galaxies containing bright AGN. Similarly, our knowledge of AGN has also benefited from this multi-wavelength approach. The fraction of galaxies hosting AGN can now be estimated from deep blank-field X-ray and optical surveys \citep[\eg][]{bar01,ste02}, providing an estimate of the duration of their active phase. 

The presence of an AGN introduces a number of complications when studying the host galaxy. This is especially true for Type~I (broad-lined) AGN, where the host galaxy can only be seen after a careful subtraction of the dominating AGN emission. This problem does not manifest itself in Type~II (narrow-lined) AGN, where the bright central emission is conveniently shielded by optically thick material. We can therefore use the AGN signatures to pinpoint the host galaxies, while still having a relatively unbiased view of the stellar population.

The downside of using type~II AGN is that their signatures are much fainter at optical wavelengths, and thus intrinsically more difficult to detect out to high redshifts. Surveys for distant type~II AGN have therefore relied on manifestations of the AGN at other wavelengths. Historically, the most successful technique has been the selection using powerful radio synchrotron emission from the AGN, which has lead to the discovery of almost 150 radio-loud type~II AGN at $z>2$ \citep{deb00b}, with the most distant known at $z=5.19$ \citep{wvb99}. This predominance of radio selection has lead to the almost unique identification of high redshift type~II AGN as 'high redshift radio galaxies' (HzRGs, $z>2$). However, recently radio-quiet type~II AGN at $z>2$ have also been found in submm \citep[][but see Vernet \& Cimatti, 2001\nocite{ver01b}]{ivi98}, X-ray \citep{stern02,nor02} and optical surveys \citep{ste02}.

Despite these new developments, radio galaxies remain amongst the best studied galaxies at high redshift, because they represent some of the most luminous stellar systems known out to $z>5$ \citep[\eg][]{bes98,deb02}. For example, deep optical spectropolarimetry can reveal a wealth of information on the stellar population of the host galaxy \citep[\eg][]{dey97}, the presence of a scattered quasar component \citep[\eg][]{ver01a}, and the kinematics, ionization mechanism and metallicity of the extended emission line regions \citep[\eg][]{vil00,deb00b}. Their emission lines (especially \Lya) are often bright enough to derive information on the density and metallicity of associated absorbers \citep{oji97a,bin00,jar02}.

The radio selection has the advantage of being insensitive to the dust properties, unlike optical selection techniques, which may miss heavily obscured systems. Such objects may be identified by their strong dust emission at far-IR wavelengths. Early observations at 800~$\mu$m and 1.2~mm by \citet{dun94} and \cite{ivi95} detected strong thermal dust emission in the radio galaxies 4C~41.17 ($z=3.8$) and 8C~1425+635 ($z=4.25$), suggesting that several HzRGs do indeed contain significant amounts of dust.
\citet{arc01} observed 47 $1<z<4.4$ radio galaxies at 850~$\mu$m using the SCUBA bolometer \citep{hol99} on the JCMT. Even after correcting for the strong negative $k-$correction effects on the Rayleigh-Jeans tail of the thermal dust spectrum \citep[\eg][]{bla93}, they found a substantial increase in the 850~$\mu$m luminosity with redshift. This is confirmed by \citet{reu03c}, who detect more than half of the sources with SCUBA in a sample of 15 radio galaxies at $z>3$. Such a redshift dependence is not seen for quasars, whose mm and submm luminosity remains almost unchanged between $z\sim 2$ and $z\sim 4$ \citep{omo03,pri03}. 
From a matched sample of $1.37 < z < 2.0$ quasars and radio galaxies selected at 151~MHz, \citet{wil02} found that the quasars are actually $\simgt 2$ times brighter at submm wavelengths than the radio galaxies. They also find an anti-correlation between the far-IR luminosity and radio size, and interpret this as a synchronization between the jet-triggering event and processes controlling the far-IR luminosity. A possible explanation for the difference between the (sub)mm redshift evolution of radio galaxies and quasars could then be that HzRGs are more massive, leading to a longer (sub)mm-luminous phase due to a larger supply of gas.

The relatively bright submm flux densities of HzRGs imply that they contain dust masses of $10^8 - 10^9$M$_{\odot}$, assuming standard conversion factors (see \S 4.2). The presence of the AGN introduces an ambiguity on the mechanism powering this dust emission: direct heating by the AGN or heating by massive stars formed in starbursts \citep[\eg][]{san96,car00,omo01,wil02}? Although the AGN could easily introduce sufficient energy to produce this emission, an increasing number of authors have argued that the starbursts are the dominating mechanism \citep[\eg][]{row00,tad02}. A strong argument in favour of the starburst origin is the detection of dust and CO emission from regions several tens of kpc away from the AGN \citep{omo96,pap00}. The high gas masses inferred from the CO detections could then be the likely reservoirs feeding these massive starbursts. The presence of CO in such reservoirs implies that they must be (at least partially) metal enriched, which is consistent with the recent detection of ionized metal lines at scales of several tens of kpc in the $z=2.922$ radio galaxy MRC~0943$-$242 \citep{vil03}. These metals may have been deposited by previous merger events, or by galaxy-scale outflows \citep[\eg][]{daw02}. 

These large dust and molecular gas masses are reminiscent of the often huge \Lya\ haloes seen around HzRGs \citep[\eg][]{oji96,ven02,vil02,reu03b}.
Moreover, $K-$band observations show that HzRGs are amongst the most massive galaxies known at each redshift \citep[\eg][]{bes98,deb02}. However, to date, the relationship between the different gas and dust components and the AGN host galaxy has remained elusive. What is the role in the galaxy formation process of the extended \Lya\ haloes and the associated absorbers seen in the rest-frame UV emission lines? Are they composed of primordial material, or have they been enriched by material from the host galaxy? Are the dust and CO reservoirs in any way linked to these?

In this paper, we report the discovery of dust and CO emission in the $z=3.087$ radio galaxy B3~J2330+3927. This is only the third $z>3$ radio galaxy where CO has been detected, after 4C~60.07 and 6C~1908+7220 \citep[$z=3.791$ and $z=3.532$;][]{pap00}, and despite several intensive searches \citep{eva96,oji97b}. Moreover, the CO emission is coincident in velocity space with associated \HI\ absorption in the \Lya\ line. This shows for the first time that these different components are indeed physically related, and trace a massive reservoir surrounding the AGN host galaxy.

Throughout this paper, we use a cosmology with a $\Lambda-$dominated Universe with H$_0 = 65h_{65}$~km~s$^{-1}$~Mpc$^{-1}$, $\Omega_{\rm M}=0.3$ and $\Omega_{\Lambda}=0.7$. At $z=3.087$, the luminosity distance is $D_L=28.34h_{65}^{-1}~{\rm Gpc} = 8.74 \times 10^{26} h_{65}^{-1}$~m, and 1\arcsec\ corresponds to 8.2~kpc. For comparison, in a flat Universe with $H_0=65h_{65}$~km~s$^{-1}$~Mpc$^{-1}$, $\Omega_{\rm M}=1.0$ and $\Omega_{\Lambda}=0.0$, $D_L=19.07h_{65}^{-1}~$~Gpc, and 1\arcsec\ corresponds to 5.5~kpc.

\section{Observations and data reduction}
\subsection{Source selection}
\citet{vig99} defined the B3-VLA sample with the aim of constructing a homogeneous spectral database for a sample of 1049 radio sources, containing flux densities in the range 151 MHz to 10.6 GHz. From this sample, we selected 104 sources with steep optical spectra ($\alpha < -1; S \propto \nu^{\alpha}$), which remained undetected on the Palomar Observatory Sky Survey. Such sources are known to be good candidates to find HzRGs \citep[\eg][]{deb00a}. We first identify the host galaxies of these candidates in sensitive optical or near-IR images, followed by optical spectroscopy to determine their redshift. In the following, we report on one of these sources, B3~J2330+3927.

\subsection{Optical and Near-IR Imaging and Spectroscopy}
\begin{figure}[ht]
\psfig{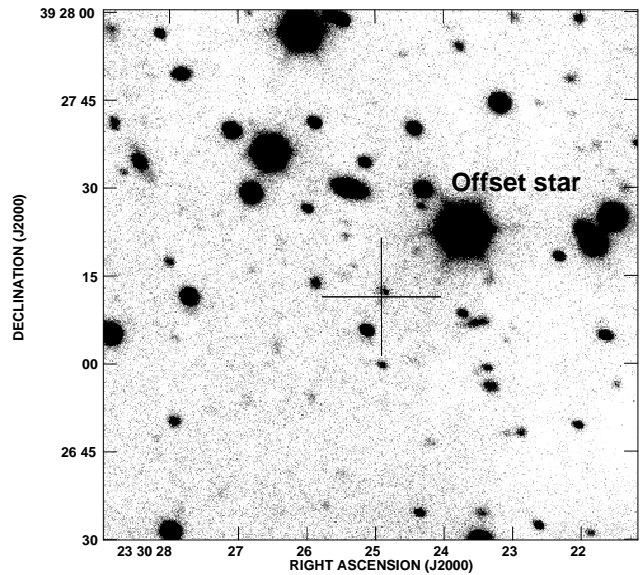}
\caption{$R-$band image of B3~J2330+3927 obtained with Keck/LRIS. The NVSS position of the radio source is marked with a cross. The coordinates of the offset star are listed in Table~\ref{B3astrophotometry}.}
\label{B3FC}
\end{figure}

On UT 1995 August 31, we obtained a 600s $R-$band image of B3~J2330+3927 with the Low Resolution Imaging Spectrograph \citep[LRIS;][]{oke95} on the Keck~I telescope. Conditions were photometric with 1\farcs0 seeing. We reduced the image using the standard techniques in the NOAO IRAF package. 
We determined the astrometric solution using 19 stars also detected in the USNO 2.0 catalog \citep{mon98}. The uncertainty in {\it relative} optical/radio astrometry is dominated by the absolute uncertainty of the optical reference frame, which is $\sim$0\farcs4 \citep[90\% confidence limit;][]{deu99}. Figure~\ref{B3FC} presents a finding chart, based on the Keck $R-$band image.

On UT 1996 September 10, we obtained a 1200s spectrum with LRIS on the Keck~I telescope. We used the 300~$\ell$~mm$^{-1}$ grating blazed at 5000\AA\ with the 1\arcsec\ wide slit resulting in a resolution of $\sim$ 10\AA\ (FWHM). The slit was oriented at PA=168\degr, and passes through objects~{\em a} and {\em b} (see Fig.~\ref{radioKoverlay}). We corrected the spectra for overscan bias, and flat-fielded them using internal lamps taken after the observations. 
We extracted a one-dimensional spectrum using a 1\arcsec\ wide aperture.
Because the conditions were not photometric, we initially flux calibrated the spectrum with an archival sensitivity function. We then determined the $R-$band magnitude from the spectrum using the IRAF task {\tt sbands}, and adjusted the flux scale to $R=24.04$, which is the 1\arcsec\ diameter aperture magnitude from our photometric $R-$band image. The main uncertainty in this procedure stems from the accuracy of the slit positioning on the peak of the $R-$band emission. We estimate the resulting spectrophotometry to be accurate up to $\sim$30\%.

On UT 1997 September 12, we obtained near-IR imaging with NIRC \citep{mat94} on the Keck~I telescope. Conditions were photometric with 0\farcs45 seeing in $K_S-$band. The total integration time of the $K_S-$image was 1140s. Observing procedures, calibration and data reduction techniques were similar to those described by \citet{deb02}. To determine the astrometric solution, we used 16 stars also detected in our LRIS $R-$band image. We estimate the {\it relative} near-IR/radio astrometric uncertainty to be $\sim$0\farcs5.

Following the $K_S-$band imaging, we also obtained near-IR spectroscopy with NIRC. We used the gr120 grism in combination with the $HK-$filter. The position angle on the sky was 302\degr, and includes only object~{\em a}. This position angle was dictated by the need for a sufficiently bright guide star in the offset guider. The dispersion was $\sim$60~\AA/pix, and the slit width was 0\farcs68, resulting in a resolution of $\sim$240~\AA\ (FWHM). We obtained 10 exposures of 180s each. We discarded two exposures which were useless due to poor seeing and/or high sky background variability. The total integration time is thus 1440s.
We reduced the spectra following the procedures described by \citet{deb98}. We extracted a one-dimensional spectrum using a 1\farcs2 wide aperture. 
Because no near-IR line-lamps were available, we used the dispersion formula listed in the NIRC manual to wavelength calibrate the spectrum. This does not allow an independent estimate of the uncertainties in the calibration, but a comparison of the 3 detected emission lines with the optical spectrum suggests it is accurate up to $\sim 70$~\AA.
To correct for atmospheric absorption lines, we divided the spectrum by the atmospheric transmission curve for Mauna Kea, produced using the program IRTRANS4\footnote{This curve is available on the UKIRT world wide web pages.}.
To flux calibrate the spectrum, we used a spectrum of the A0 star HD220750, which was obtained immediately subsequent to our B3~J2330+3927 observations. We reduced this spectrum in the same way as B3~J2330+3927. To construct a sensitivity function, we assumed HD220750 has a blackbody spectrum with $T_{\rm eff}=$9850~K implying $V-K=-0.03$ \citep{jon66}. Using $V=7.01$ from the SIMBAD database, this provides a first order flux calibration. To refine the absolute flux calibration, we determined the $K_S-$band magnitude from the spectrum using the IRAF task {\tt sbands}, and adjusted the flux scale to $K=19.18$, which is the 1\arcsec\ diameter aperture magnitude from our photometric $K_S-$band image.

We corrected both the optical and near-IR spectra for Galactic reddening using the \cite{car89} extinction curve with $A_V=0.39$, as determined from the dust maps of \citet{schl98}.

\subsection{VLA radio imaging}
\begin{figure}[ht]
\psfig{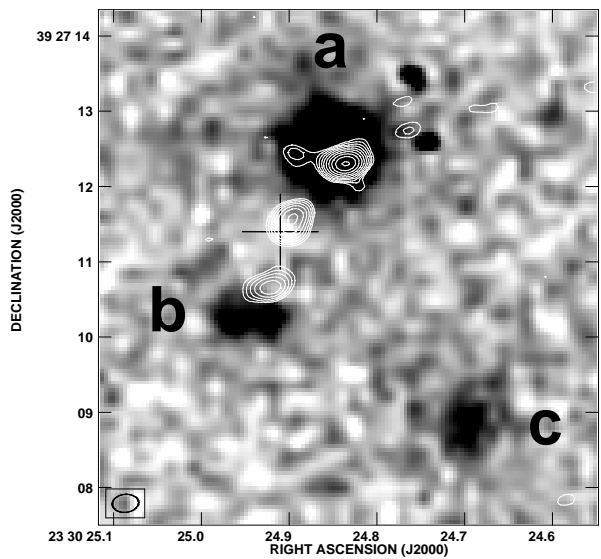}
\caption{8.46~GHz VLA map of B3~J2330+3927 overlaid on our Keck $K_S-$band image (subject to a 0\farcs5 uncertainty in the registration). We identify 3 components {\em a}, {\em b}, and {\em c}, which may be part of the same physical system (the 2 small peaks west and north-west of object~{\em a} are bad pixels). The cross marks the position in the NVSS catalogue, which is within 0\farcs2 from the position found in the 1.4~GHz A-array data of \citet{vig99}. The FWHM of the synthesized 8.46~GHz VLA beam is shown in the bottom left corner.}
\label{radioKoverlay}
\end{figure}

To determine the morphology of the radio source, we observed B3~J2330+3927 for 1 hour with the VLA \citep{nap83} on UT 2002 March 30. We used the A-array at a frequency of 8.46~GHz and a bandwidth of 50~MHz, resulting in a resolution of $0\farcs35 \times 0\farcs24$. Due to technical problems, no primary flux calibrator was observed. We used an observation of 3C~286 from the next day to calibrate the data. However, the absolute flux calibration cannot be considered reliable, and we therefore only use this observation to determine the source morphology. We used the standard data reduction techniques in the NRAO \aips\ package. Because the source is too faint, we did not self-calibrate the data. Figure~\ref{radioKoverlay} shows the resulting 8.46~GHz image.

\subsection{IRAM interferometric Observations}
B3~J2330+3927 is one of the brightest 850~$\mu$m sources detected in the SCUBA survey of $z>3$ radio galaxies by \citet{reu03c}. Because gas and dust emission are often related, we selected B3~J2330+3927 as a prime target to search for redshifted CO emission in HzRGs.

\begin{figure}[ht]
\psfig{file=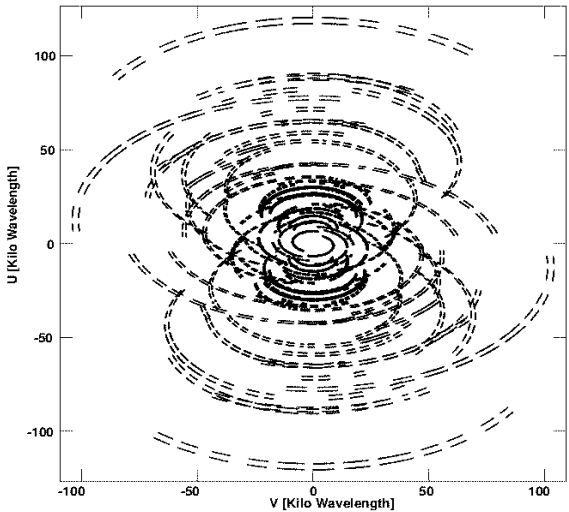,width=8cm,angle=-90}
\caption{The $u-v$ plane coverage for B3~J2330+3927 observed at 113~GHz with the IRAM Plateau de Bure Interferometer. Note the excellent coverage obtained with 3 different configurations using 4 to 6 antennae.}
\label{B3UV}
\end{figure}

We observed B3~J2330+3927 with the IRAM Plateau de Bure Interferometer during 8 individual sessions between 2001 May and 2002 February. We used 3 different configurations (D, C and B) with 4 to 6 antennae, depending on availability. The total on-source observing time was $\sim$27~hours (excluding calibrations), with an excellent sampling of the uv-plane (Fig.~\ref{B3UV}). The resulting 112.798~GHz synthesized beam size is $3\farcs2 \times 2\farcs3$ at position angle 56\degr.
The correlator setup for the 3~mm receivers involved $4 \times 160$~MHz modules covering a total bandwidth of 560~MHz tuned in single sideband (SSB) mode.

The central frequency was originally tuned to 112.834~GHz to observe CO~$(J=4-3)$ ($\nu_{\rm rest}=461.040$~GHz) at $z=3.086$, the redshift of the \Lya\ line. However, after an initial assessment of the data from the first session, we re-centered the central frequency to 112.780~GHz ($z=3.092$) to better cover the full extent of the detected emission line, which appeared to be offset from the optical redshift (see below).

We simultaneously observed at 230.538~GHz~(1.3~mm) in double side-band (DSB) mode to probe the spatial extend of the thermal dust emission. However, in the following, we consider only those 1.3~mm data that were obtained when the atmospheric water vapour content was $<3$~mm. The total 1.3~mm integration time is therefore only 17.7~h. The synthesized beam size at 1.3~mm is $1\farcs9 \times 1\farcs5$.

After phase and amplitude calibration, we combined the data with different central frequencies into an expanded data-set of 271 spectral channels of 2.5~MHz each, calculating the proper statistical weight of each channel individually.
The outer channels of the spectral cube therefore have larger uncertainties (see Fig.~\ref{B3COposa}). 

\subsection{IRAM Single Dish Bolometer Observations}
To determine the total 1.3~mm flux density, B3~J2330+3927 was observed in service mode with the 117-channel Max Planck  Bolometer array \citep[MAMBO;][]{kre98} at the IRAM 30m telescope on Pico Veleta, Spain. The effective bandwidth centre for steep thermal spectra is $\sim$250~GHz. We observed the source with the array's central channel, using the standard 'ON-OFF' observing sequence, in which the secondary mirror chops in azimuth by 32\arcsec\ at a rate of 2~Hz.  We adopted a calibration factor of 35000 counts per Jansky, which we estimate is accurate to within 20\%.

The observations were split in 2 sessions: 5$\times$16 subscans of 12s on UT 2002 October 30 and 8$\times$16 subscans on UT 2002 November 2, resulting in a total on-source integration time of 21~minutes.
We reduced the data with the MOPSI software package \citep{zyl98}, including subtraction of the correlated sky noise. This yields a target flux of $S_{250\rm GHz}=4.8\pm1.2$~mJy.

\subsection{WSRT 21~cm line Observations}
We observed B3~J2330+3927 with the Westerbork Synthesis Radio Telescope (WSRT) on several occasions between UT 2001 July 1 and 2002 September 10 to search for redshifted \HI\ 21~cm (1420.4057~MHz) absorption against the synchrotron emission from the radio source.
Unfortunately, a significant fraction of these observations were rendered useless due to external radio frequency interference (RFI), or by an unstable ionosphere.
In the following, we shall therefore only use the highest quality observations, with a total integration time of 22 hours.

We used the DZB backend with a 10~MHz bandwidth centered on 347~MHz and 128 channels giving a velocity resolution of $\sim 16$\kms\ before Hanning smoothing the data.

We made a line cube using uniform weighting, giving a resolution of $61\arcsec \times 36\arcsec$ at PA=0 (north-south).  The noise in the single channel is $\sim 1.7$~mJy/Beam/channel (after Hanning smoothing). 
We have also obtained a continuum image of the field.
At the observed frequency and resolution of the WSRT observations the continuum emission of B3~J2330+3927 is unresolved with a peak flux of 405~mJy, consistent with the 325~MHz WENSS and 365~MHz Texas flux densities (see also Fig.~\ref{B3SED}).

\section{Results}

\subsection{Identification and spectroscopy of the optical/near-IR components}

Figure~\ref{radioKoverlay} shows the location of the radio source on the $K_S-$ band image. The radio source, consisting of 3 components with a largest separation of $1\farcs9$, is located exactly in between two objects detected in both the $R-$ and $K_S-$band images. We also detect a resolved component in the $K_S-$band image 3\farcs5 southwest of the radio source, which may be related. Table~\ref{B3astrophotometry} lists the positions and photometry of these 3 components, which we label as {\em a}, {\em b}, and {\em c}. 

\begin{figure}[ht]
\psfig{file=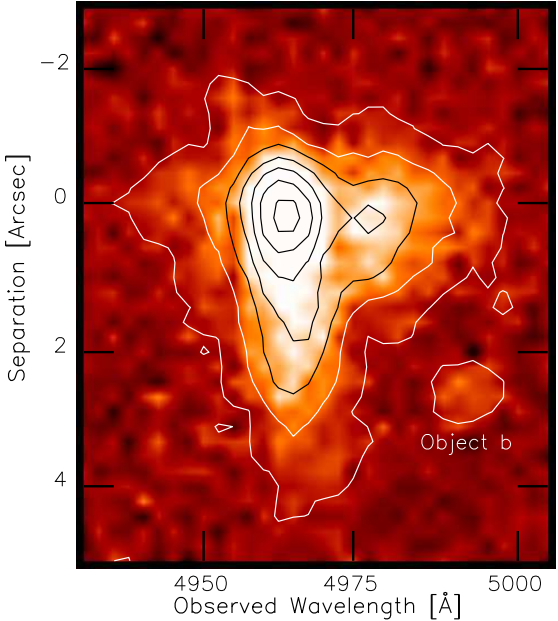,width=12cm}
\caption{Two-dimensional spectrum, centered on \Lya. Object~{\em b} is a companion object, offset by 1500\kms with respect to the AGN. The y-axis has been reversed to facilitate comparison with Figure~\ref{radioKoverlay}. Object~{\em a} is at the origin of the y-axis.}
\label{B3Lya2D}
\end{figure}

\begin{figure}[ht]
\psfig{file=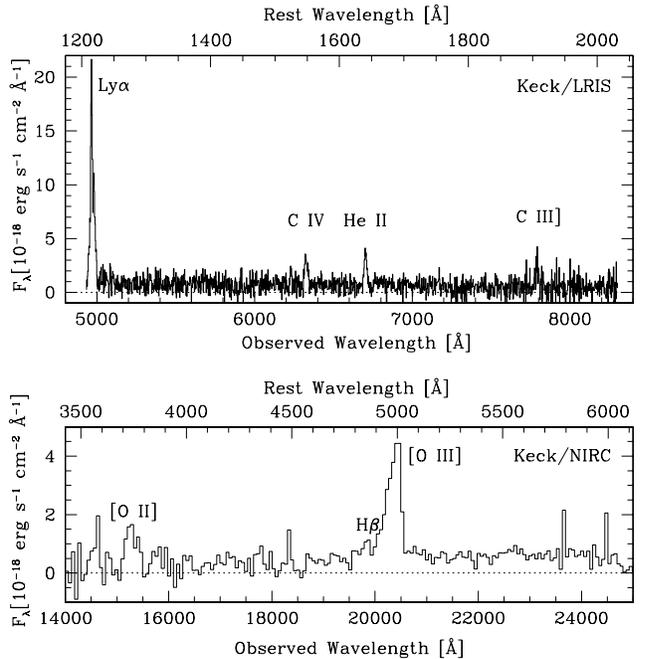,width=9cm}
\caption{Optical and near-IR spectra of object~{\em a}, extracted with $1\arcsec \times 1\arcsec$ and $0\farcs68 \times 1\farcs2$ apertures, respectively.}
\label{B3spectra}
\end{figure}

The optical spectroscopic slit included objects~{\em a} and {\em b}, while the near-IR slit included only object~{\em a}. Figures~\ref{B3Lya2D} and \ref{B3spectra} show that object~{\em a} has the classical spectrum of a type~II AGN, showing narrow emission lines and a faint continuum emission. 
We only detect a single, weak emission line in object~{\em b} (Fig.~\ref{B3Lya2D}), which can only be interpreted as \Lya\ at $z=3.107$; we can exclude the alternative identification as \OII\ at $z=0.339$, because in this case, we would have also detected the \OIII\ line. At $z=3.107$, the \OIII\ line may contribute significantly to the $K-$band flux, but object~{\em b} cannot be a reflection nebula, as emission is also detected in the $R-$band, which is free of strong emission lines.
Object~{\em b} therefore appears to be a companion object, offset by +1500\kms with respect to the systemic redshift of object~{\em a}. 

\begin{table}
\caption{Astrometry and photometry (in 2\arcsec\ apertures) of the optical/near-IR components of B3~J2330+3927.}
\label{B3astrophotometry}
{\scriptsize
\begin{tabular}{rrrrr}
\hline
Object & RA(J2000) & DEC(J2000) & $R$(mag) & $K_S$(mag) \\
\hline
{\em a} & $23^h 30^m 24\fs85$ & $39\degr 27\arcmin 12\farcs6$ & 23.08$\pm$0.12 & 18.79$\pm$0.03 \\
{\em b} & $23^h 30^m 24\fs94$ & $39\degr 27\arcmin 10\farcs3$ & 24.18$\pm$0.22 & 20.76$\pm$0.11 \\ 
{\em c} & $23^h 30^m 24\fs70$ & $39\degr 27\arcmin 08\farcs9$ & $>$25          & 20.70$\pm$0.10 \\
Offset & $23^h30^m23\fs71$ & $39\degr 27\arcmin 22\farcs8$ & 15.3 & 14.4$\pm$0.2 \\
\hline
CO peak & $23^h30^m24\fs84$ & $39\degr 27\arcmin 12\farcs2$ & & \\
\hline
\end{tabular}
}
\end{table}

Figure~\ref{B3spectra} shows the 1-dimensional optical and near-IR spectra of B3~J2330+3927 (only object~{\em a} is included in the apertures). We detect 4 emission lines in the optical spectrum (see Table~\ref{B3lines}), and 3 emission lines in the near-IR spectrum. Because the low resolution and uncertain wavelength calibration of the near-IR spectrum, we only use the optical spectrum to determine the systemic redshift of B3~J2330+3927. We adopt the redshift of the \HeII\ line, $z=3.087 \pm 0.004$, because this is the brighest non-resonant line.

\begin{table*}[ht]
\caption{Emission line measurements for object~{\em a}}
\label{B3lines}
\begin{tabular}{llrrrr}
\hline
Line & $\lambda_{\rm obs}$ & $z_{\rm em}$ & $10^{16}\times F_{\rm int}$ & $\Delta V_{\rm FWHM}^a$ & $W_{\lambda}^{\rm rest}$ \\
 & \AA & &  erg s$^{-1}$ cm$^{-2}$ & km s$^{-1}$ & \AA \\
\hline
\Lya  & 4968.3$\pm$0.7 & 3.086 & 4.4$\pm$0.4   & 1500$\pm$200 & 120$\pm$20\\
\CIV  & 6326.3$\pm$3.0 & 3.084 & 0.44$\pm$0.06 &  800$\pm$300 &  14$\pm$ 3\\
\HeII & 6703.3$\pm$2.0 & 3.087 & 0.56$\pm$0.07 &  700$\pm$225 &  20$\pm$ 4\\
\CIII & 7791.6$\pm$3.5 & 3.082 & 0.47$\pm$0.07 &  700$\pm$300 &  24$\pm$ 8\\
\OII  & 15280$\pm$100  & 3.10  & 3.9$\pm$0.3   & ...          & 250$\pm$150\\
\Hbeta& 19840$\pm$100  & 3.08  & 0.7$\pm$0.2   & ...          &  27$\pm$15\\
\OIII & 20370$\pm$100  & 3.07  &11.5$\pm$1.0   & ...          & 370$\pm$80\\
\hline
\end{tabular}

$^a$ Deconvolved with the instrumental resolution.
\end{table*}

The HK spectrum allows us to determine the contribution from emission lines to the $K-$band magnitude of object~{\em a}. We find that the \OIII\ line contributes 36\% (0.5 magnitudes) to the $K_S-$band flux (the \Hbeta\ line is much weaker and can be safely ignored). This value is comparable to the 45\% contribution in the $z=3.594$ radio galaxy MG~J2144+1928 \citep{arm98}, and to the 34~\% average contribution in a sample four radio galaxies where the \OIII\ line falls in the $K-$band \citep{eal93}. After the emission line contribution is removed, the 64~kpc $K-$band magnitude ($K=18.79 \pm0.03$) is consistent with those for powerful HzRGs at this redshift \citep{wvb98}. This shows that the inclusion of the strong \OIII\ line can increase the scatter in the $K-z$ diagram, but it is unlikely to be a dominant factor \citep[see also][]{jar01}.

\subsection{Where is the AGN?}
Our spectroscopic results identify object~{\em a} as the most likely position of the AGN. This appears contradictory to the 8.4~GHz radio morphology shown in Figure~\ref{radioKoverlay}, and summarized in Table~\ref{B3radio}. The most straightforward interpretation of the VLA map is a core surrounded by 2 hotspots, as seen in many HzRGs \cite[\eg][]{car97,pen00}. The central component appears marginally resolved with our $0\farcs35 \times 0\farcs24$ resolution, which looks like a carbon-copy of the $z=4.25$ radio galaxy 8C~1435+635 \citep{lac94}. The radio morphology therefore suggests that the central component marks the position of the AGN. Unfortunately, we do not have spectral index information on the individual radio components to examine this interpretation in more detail.

\begin{table}
\caption{Radio components of B3~J2330+3927. The 8.4~GHz fluxes are normalized the northern component.}
\label{B3radio}
{\scriptsize
\begin{tabular}{rrrr}
\hline
Component & RA(J2000) & DEC(J2000) & Normalized flux \\
\hline
NVSS   & $23^h30^m24\fs91$ & $39\degr 27\arcmin 11\farcs4$ & --- \\
North  & $23^h30^m24\fs836$ & $39\degr 27\arcmin 12\farcs31$ & 1.00 \\
Centre & $23^h30^m24\fs897$ & $39\degr 27\arcmin 11\farcs54$ & 0.43 \\
South  & $23^h30^m24\fs922$ & $39\degr 27\arcmin 10\farcs67$ & 0.23 \\
\hline
\end{tabular}
}
\end{table}

Given the $0\farcs4$ astrometric uncertainty, we can state that the central radio component does not coincide with object~{\em a}. If the nominal astrometry in Fig.~\ref{radioKoverlay} is correct, then the northern and southern radio lobes coincide with objects~{\em a} and {\em b}, respectively. 
We can also exclude the presence of a significant radio component north of object~{\em a} because the positions of the low-resolution radio data from the NVSS and from the VLA maps of \citet{vig99} (cross in Figure~\ref{radioKoverlay}) are within 0\farcs2 of the central radio component in our 8.4~GHz map.

We propose two possible interpretations of the observed radio/near-IR configuration. First, the northern component may be the radio core of an extremely asymmetrical source, or a one-sided jet, much like the $z=4.1$ radio galaxy TN~J1338$-$1942 \citep{deb99}. \citet{mcc91} have argued that the extended optical emission lines are brightest on the side of the closer of the two radio lobes, which would be consistent with our spectroscopy (Fig.~\ref{B3Lya2D}).
The slight mis-alignment of the 3 components may then indicate a precession of the radio jet, which is often seen in FR~II sources \citep[\eg][]{den02}.

The alternative interpretation is that the AGN is located at the position of the central radio component, but is either heavily obscured by dust \citep[\eg][]{reu03a}, or not located at the centre of the host galaxy \citep[\eg][]{qui01}. A potential problem with this interpretation could be the ionization of gas at object~{\em a}, 10~kpc away from the AGN. However, spatially extended \CIV\ and \HeII\ emission has been seen in several HzRGs \citep[\eg][]{ove01,max02}.

\subsection{Possible synchrotron contribution at mm wavelengths}
Because B3~J2330+3927 is a powerful radio galaxy, it is important to estimate the possible contribution from synchrotron emission in the observed mm and sub-mm bands. We used the multi-frequency data of \citet{vig99} to determine the radio to submm spectral energy distribution. We also added the 365~MHz flux density from the Texas survey \citep{dou96}, our 250~GHz MAMBO detection (\S 2.5), the SCUBA 353~GHz (850~$\mu$m) detection of \citet{reu03c}, and the 3mm (112.7~GHz) continuum emission seen in our PdBI data (see next section). We also include our tentative 1.25~mm (230.5~GHz) detection, although we suspect that this emission has been over-resolved (see \S3.5).

\begin{figure}[ht]
\psfig{file=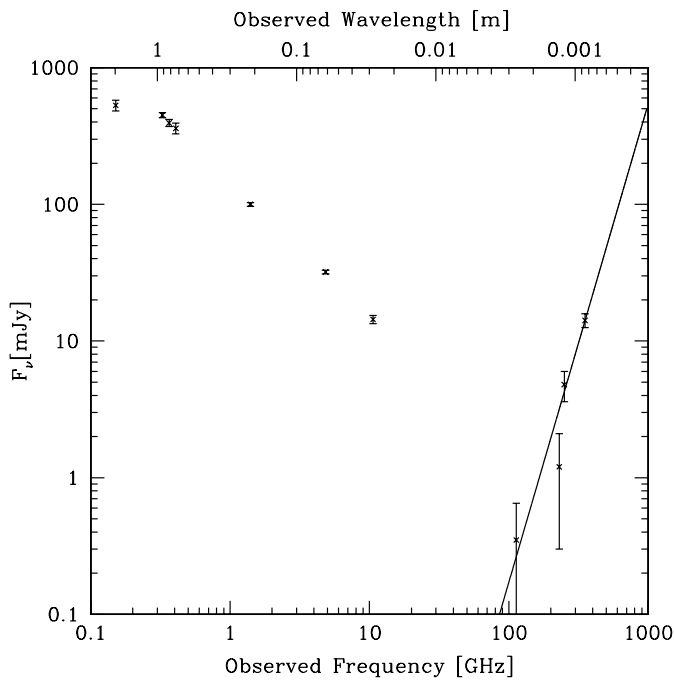,width=9cm}
\caption{Radio to submm spectral energy distribution of B3~J2330+3927. The solid line is a modified black body curve with $T=50$~K and $\beta=1.5$, normalized at 352~GHz. The 230.538~GHz PdBI measurement severely over-resolves the thermal dust emission.}
\label{B3SED}
\end{figure}

Figure~\ref{B3SED} shows this combined spectral energy distribution (SED). It is obvious that a synchrotron contribution would be negligible at 353~GHz (850~$\mu$m), but may contribute to the 113~GHz and 230~GHz detections. A constant power-law extrapolation of the synchrotron spectrum from the 10.6~GHz data would predict a synchrotron flux density of 1.3~mJy at 113~GHz and 0.6~mJy at 230.5~GHz. However, the synchrotron spectra of powerful radio galaxies often steepen at high frequencies \citep{arc01,and02}, with spectral steepenings up to $\Delta \alpha \sim 0.5$ \citep{mur99}. This suggests the constant power-law extrapolation should be considered as an upper limit. 
To estimate the 112.7~GHz contribution more accurately, we have fit the radio SED, using the Continuous Injection model of \citet{pac70} and the Single Injection model of \citet{jaf73}. Even with the latter model (which leads to an exponential steepening of the spectrum at high frequency), we found that the minimum flux density predicted for the 112.7~GHz point is $\sim 0.6-0.8$~mJy. More observations between 10.6~GHz and 113~GHz would be needed to predict this contribution more accurately.

Note that these values refer to the integrated flux density of the radio source. However, Figure~\ref{radioKoverlay} shows that the radio source is resolved into three components, which appear to have roughly similar flux densities at 8.4~GHz. Unfortunately, we do not have spectral index information of the individual components, but in most radio galaxies, the outer lobes have steeper spectra than the core \citep{car97,pen00}, although sources with relatively steep spectrum cores have also been seen at high redshift \citep{ath97}. The orientation of the 112.7~GHz PdBI synthesized beam is roughly perpendicular to the radio axis, which means we should slightly resolve the synchrotron component at 112.7~GHz. Depending on the spectral index of the northern hotspot, this further reduces the synchrotron contribution at the position of the AGN to $\simlt 0.5$~mJy. 

\subsection{Identification of the CO emission}

\begin{figure}[ht]
\psfig{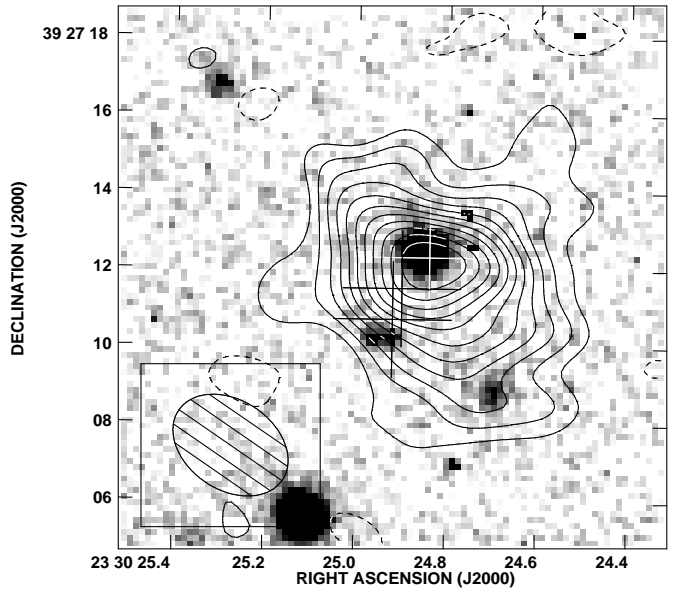}
\caption{
Integrated CO~$J=4-3$ emission towards B3~J2330+3927 overlaid on the Keck/NIRC $K-$band image. Contour levels start at 2~$\sigma$, and increase by 1~$\sigma$, with $\sigma=0.2$~mJy~Beam$^{-1}$ (negative contours are dotted). The 3 crosses mark the positions of the radio components in Fig.~\ref{radioKoverlay}. The FWHM of the synthesized beam is shown in the bottom left corner.}
\label{COoverlay}
\end{figure}

Figure~\ref{COoverlay} shows the 3mm PdBI data, integrated between 112.515~GHz and 112.723~GHz, which represents the full width of the redshifted CO~$J=4-3$ line (see below). The emission clearly peaks at the position of object~{\em a}, which is the most likely position of the AGN (see \S~3.2). The emission does not follow the extension of the radio source, confirming that a possible synchrotron contribution would be minimal.

\begin{figure}[ht]
\psfig{file=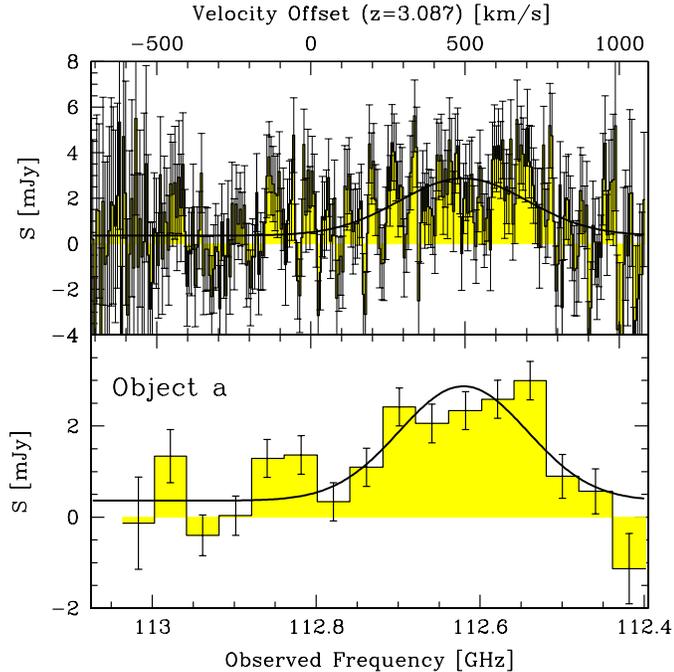,width=9cm}
\caption{CO~$J=4-3$ spectrum of B3~J2330+3927 extracted at object~{\em a}. The top panel shows the unbinned spectrum, while the bottom panel has been re-binned with $\Delta \nu_{\rm chan}=40$~MHz (106.5\kms). The velocity scale is with respect to the central frequency of CO~$J=4-3$ at $z=3.087$. The solid line is a Gaussian fit with an underlying continuum emission.}
\label{B3COposa}
\end{figure}

Figure~\ref{B3COposa} shows the PdBI CO~$J=4-3$ spectrum of B3~J2330+3927 extracted at the position of object~{\em a}. We show the unbinned spectrum, and a spectrum binned to 105~\kms. A broad emission line is clearly seen around 112.6~GHz, with a weak underlying continuum emission.
We fit a Gaussian profile with an underlying continuum emission to determine the line and continuum parameters. Table~\ref{3mmfit} gives the fit parameters for 5 different bin sizes used. As long as the bin size does not exceed 40~MHz, the derived parameters are not very sensitive to the bin widths.

In the previous section, we estimated the contribution from synchrotron emission to be $<0.5$~mJy, so the broad, $\sim$2~mJy line appears securely detected. We interpret this emission line as CO~$J=4-3$ at $z=3.094$, \ie\ offset by $\sim$500~\kms\ to the red from the optical redshift, determined from the \HeII\ line. Such offsets have also been seen in high S/N CO identifications of high redshift quasars \citep[\eg][]{gui99,cox02a}.
The $\sim$0.3~mJy continuum emission seen in the 1.3~mm spectrum (Fig.~\ref{B3COposa}) is consistent with the extrapolation of the thermal dust spectrum (Fig.~\ref{B3SED}), but it is also likely that there is a synchrotron contribution from the northern radio component, especially if the latter would have a flatter spectral index than the integrated radio emission.

The total velocity width of the CO-line is $\sim$500\kms (see Fig.~\ref{B3COposa}), and the integrated flux over these five 106.5~\kms\ channels is $S_{\rm CO} \Delta V = 1.3 \pm 0.3$~Jy~\kms. Note that this value has been corrected for the underlying continuum emission ($\sim$ 0.3~mJy). This represents one of the faintest un-lensed detections of CO emission to date.

To determine if the CO emission in spatially resolved, we have subtracted different source models (point source, circular or elliptical Gaussians) from the UV-data. We find no clear evidence that the integrated CO emission is spatially resolved, despite the suggestion in Fig~\ref{COoverlay} that the emission appears resolved towards object~{\em c}. The latter may be due to the low S/N of the outer contours. 
Note that due to our limited frequency coverage, we may have missed CO emission from components with a large velocity offset. For example, object~{\em b} is known to have a velocity offset of 1500~\kms (Fig.~\ref{B3Lya2D}), so any CO emission from this object would not have been detected in our observations. Deeper integrations with a wider frequency coverage would be needed to examine the spatial structure of the CO emission.

\begin{table}
\caption{Gaussian fitting coefficients of the CO line in object~{\em a}}
\label{3mmfit}
\small
\begin{tabular}{rrrrrr}
\hline
Bin width & \# bins & $S_{\rm cont}$ & $S_{\rm peak}$ & $\nu_{\rm centre}$ & $\Delta V_{\rm FWHM}$  \\
MHz & & mJy & mJy & GHz & km/s \\
\hline
 2.5 & 270 & 0.36 & 2.51 & 112.620 & 207 \\
10.0 &  67 & 0.33 & 2.55 & 112.620 & 211 \\
20.0 &  33 & 0.36 & 2.47 & 112.623 & 208 \\
40.0 &  16 & 0.28 & 2.51 & 112.621 & 223 \\
80.0 &   8 & 0.52 & 2.22 & 112.614 & 208 \\
\hline
\end{tabular}
\normalsize
\end{table}

\subsection{Identification of the spatially extended thermal dust emission}
From the radio to submm spectral energy distribution (Fig.~\ref{B3SED}), it is obvious that the interferometer data under-estimate the dust emission at 230~GHz.
We now examine this possibility in detail using our 230.5~GHz PdBI map (Fig.~\ref{B3K1.3mm}).

Extrapolating the 353~GHz (850~$\mu$m) SCUBA detection (14.15 $\pm$ 1.65~mJy) using a blackbody with $T=50~K$ predicts an integrated dust emission flux density of 4.2~mJy at 230~GHz and 0.45~mJy at 113~GHz. Figure~\ref{B3K1.3mm} shows the central part of our 230.5~GHz PdBI map. We find a marginal detection of dust emission with an integrated flux density of 1.2$\pm$0.9~mJy.
This appears inconsistent with the SCUBA detection. Such a discrepancy has also been seen in high redshift quasars, where the PdB interferometer data underestimate the flux by as much as an order of magnitude when compared to IRAM 30m single dish measurements at the same frequency \citep{gui99,omo01}. The most likely explanation is that the dust emission is extended with respect to the much smaller interferometer synthesized beam. 

Although our 230~GHz detection is marginal at most, it is concentrated near the peak of the CO emission and the most likely position of the AGN (see \S 3.2), supporting the reality of this detection. We conclude that there is evidence for a two-component spatial distribution of the dust emission: first, a compact component, which may be heated by the AGN, and second, a component significantly larger than the $1\farcs9 \times 1\farcs5$ PdBI synthesized beam at 230~GHz.

\begin{figure}[ht]
\psfig{file=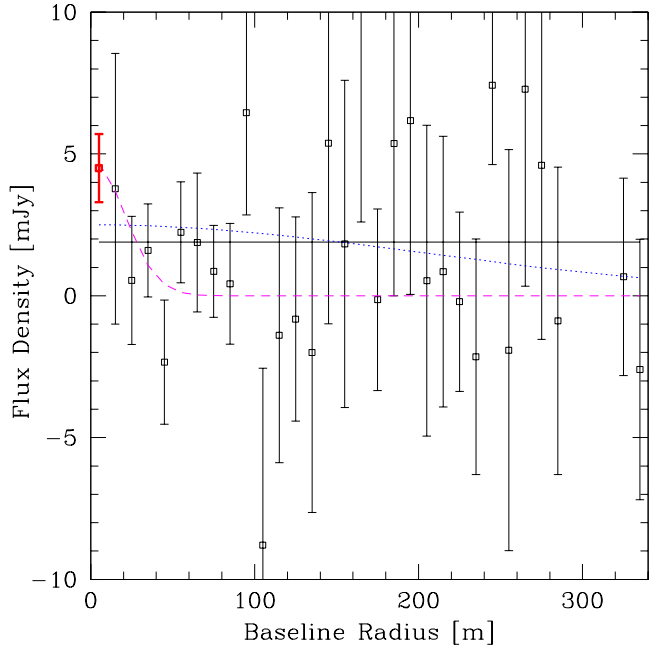,width=9cm}
\caption{Visibility amplitude versus baseline radius in the the uv-plane for the 230~GHz PdB interferometer data (black/thin points), and the MAMBO/IRAM single dish data (thick/red point) interpolated to 230~GHz. The solid line represents a point source of 1.8~mJy, the dotted/blue and dashed/magenta lines a circular Gaussian source with a FWHM of 0\farcs5 and 5\arcsec, respectively.}
\label{uv230GHz}
\end{figure}

The fact that the $\sim$0.3~mJy continuum emission seen in the 113~GHz spectrum (Fig.~\ref{B3COposa}) is close to the extra\-polation of the thermal dust emission (despite a possible synchrotron contribution) suggests that the $3\farcs2 \times 2\farcs3$ synthesized beam does not over-resolve this dust emission. In theory, we could determine the scale size of the thermal dust emission by combining the zero-spacing MAMBO flux density (scaled to 230~GHz) with the PdB interferometer data (Fig.~\ref{uv230GHz}). However, due to the low signal/noise of our data, we can only constrain the scale size between 0\farcs5 and 5\arcsec.

\begin{figure}[ht]
\psfig{file=ms3003.f9.ps,width=9cm,angle=-90}
\caption{IRAM PdBI 230.5~GHz (1.3~mm) map of B3~J2330+3927 overlaid on the Keck/NIRC $K-$band image. Contour levels start at 1~$\sigma$, and increase by 1~$\sigma$, with $\sigma=0.45$~mJy~Beam$^{-1}$ (negative contours are dotted). The 3 crosses mark the positions of the radio components in Fig.~\ref{radioKoverlay}. The FWHM of the synthesized beam is shown in the bottom left corner.}
\label{B3K1.3mm}
\end{figure}

\subsection{No \HI\ 21~cm absorption}

Despite our deep and repeated WSRT observations, we have not been able to convincingly detect \HI\ 21~cm absorption against the radio source. This is consistent with unsuccessful WSRT searches in 6 HzRGs \citep{rot99}.
To date, \HI\ 21~cm absorption has been solidly detected in only one HzRG, viz.\ B2~0902+34 \citep[$z=3.3909$][]{uson91,bri93}. 
This raises the question as to why B2~0902+34 stands out. One clue may come from its bizarre radio morphology \citep{car95}. \citet{rot99} argue that the \HI\ is located in a kpc-sized disc, which can be seen in absorption due to a fortuitous orientation of the radio source. They further suggest that a high core fraction may characterize such orientations: 15\% for B2~0902+34, compared to 0.6\% and 2.9\% for the 6 HzRGs where no 21~cm absorption has been detected. However, in B3~J2330+3927, we find a core fraction of 60\% or 26\% (see Table~\ref{B3radio}), depending if we adopt the northern or central radio component as the core (see \S 3.2). We conclude that the present sample of 8 HzRGs where \HI\ 21~cm absorption has been searched is clearly too small to address the origin of the absorbing neutral gas.

\section{Mass estimates}

The detection of the different emission and absorption components in B3~J2330+3927 suggests it is a massive system. In this section, we use the above data to derive 5 independent mass estimates of the gas and dust components.
We discuss each component separately, and summarize our results in Table~\ref{massestimates}.
We note that several high redshift CO detections have been helped by gravitational lensing \citep[for reviews, see][]{com99, cox02b}. However, from our radio and optical/near-IR imaging, we find no indications that B3~J2330+3927 could be amplified by gravitational lensing, and we do not consider this in the following. 

\subsection{Molecular gas mass}
We can estimate the molecular gas content from our CO line detection by making an assumption for the standard conversion factor $X_{\rm CO}=M({\rm H_2})/L^{\prime}_{\rm CO}(1-0)$, where 
\begin{equation}
L^{\prime}_{\rm CO} = \Big(\frac{c^2}{2k}\Big)S_{\rm CO}\Delta V \nu_{\rm obs}^{-2}D_L^2(1+z)^{-3}
\end{equation}
is the CO line luminosity measured in K~\kms~pc$^2$ \citep{sol92}. The standard value for Galactic molecular clouds is $X_{\rm CO}=4$M$_{\odot} ({\rm K} \kms {\rm pc}^2)^{-1}$. However, there are strong indications from studies of extreme starbursts in ultra luminous infra-red galaxies (ULIRGs) that this value should be considerably smaller for dense starbursts and AGN. We shall therefore assume $X_{\rm CO} \approx 0.8$~M$_{\odot} ({\rm K} \kms {\rm pc}^2)^{-1}$, as given by \citet{dow98}.

Because we have observed the CO~$J=4-3$ transition, we also need to estimate the $(4-3)/(1-0)$ line ratio $r_{43}$. \citet{pap00} have used a large velocity gradient code to predict the $r_{43}$ ratio in two radio galaxies at similar redshifts as B3~J2330+3927. Using an interstellar matter environment similar to that of a local average starburst, they predict $r_{43}=0.45$. 

Following \citet{sol92}, we find $L^{\prime}_{\rm CO}(4-3)=3.9 \times 10^{10} h_{65}^{-2}$~K~\kms~pc$^2$. Inserting the values given above, we find $$M({\rm H_2}) = \Bigg(\frac{X_{\rm CO}}{r_{34}}\Bigg) L^{\prime}_{\rm CO}(4-3) \approx 7h_{65}^{-2} \times 10^{10}M_{\odot}.$$ 
This value is slightly lower than the ones found in the previous two HzRGs, 4C~60.07 and 6C~J1908+7220 (2.4 and $1.35 \times 10^{11}\rm M_{\odot}$, respectively, in our adopted cosmology).

\subsection{Dynamical mass from CO}
Our PdBI CO data does not have sufficient S/N to derive detailed spatial and spectral information needed to fit models of rotating disks, such as those often seen in local ULIRGs and recently also at high redshift \citep[$z=2.8$; ][]{gen03}. However, we can derive a rough upper limit on the dynamical mass by assuming that the size of a possible rotating disk is smaller than the resolution of our PdBI 113~GHz observations. Provided that the CO is effectively located in such a disk, this is probably close to the actual extent, given our estimated scale size of the dust emission, which is often correlated with the CO emission.

Following \citet{pap00}, the dynamical mass $M_{\rm dyn}$ is then given by  
\begin{equation}
M_{\rm dyn} \approx 1.16\times 10^9  \Bigg(\frac{\Delta V_{\rm FWHM}}{100 \kms}\Bigg)^2 \Bigg(\frac{L}{\rm kpc}\Bigg) (\sin^2 i)^{-1}~{\rm M_{\odot}},
\end{equation} 
where $i$ is the inclination of the disk. Adopting $L<25$~kpc and $\Delta V_{\rm FWHM}=220$~\kms, we find $$M_{\rm dyn}<6.4\times 10^{10}(\sin^2 i)^{-1} \rm M_{\odot}.$$ Although our estimates of $M({\rm H_2})$ and $M_{\rm dyn}$ are very uncertain, this suggests that a substantial fraction of the total mass is in the form of molecular gas.

\subsection{Dust mass}
Figure~\ref{B3SED} shows that the 850~$\mu$m flux of B3~J2330+3927 is almost completely dominated by thermal dust emission. We can therefore estimate the dust mass $M_{\rm d}$ using: 
\begin{equation}
M_{\rm d} = \frac{S_{\rm obs} D_L^2}{(1+z) \kappa_{\rm d}(\nu_{\rm rest}) B(\nu_{\rm rest},T_{\rm d})},
\end{equation} 
with $S_{\rm obs}$ the observed flux density at a wavelength where the SED is dominated by thermal dust emission, $\kappa_{\rm d}(\nu_{\rm rest})$ the rest-frequency mass absorption coefficient, $B(\nu_{\rm rest},T_{\rm d})$ the Planck function for isothermal emission from dust grains radiating at a temperature $T_{\rm d}$, and $D_L$ the luminosity distance. 

We need to make some assumptions for $T_{\rm d}$ and $\kappa_{\rm d}(\nu_{\rm rest})$. Because we observe the steep Rayleigh-Jeans tail of the dust distribution, a small change in the adopted dust temperature may cause a significant change in the derived dust mass \citep[see \eg][]{hug97,mcm99}. 
The uncertainties in both the SCUBA 850~$\mu$m and the MAMBO 1.3~mm flux densities cannot constrain $T_{\rm d}$ tightly (see Fig.~\ref{B3SED}). We will therefore adopt $T_{\rm d}=50$~K, which has been used for most other high redshift AGN \citep[\eg][]{pap00,arc01,omo01}, and provides a good fit to the thermal part of the SED of B3~J2330+3927.

The rest-frame mass absorption coefficient $\kappa_{\rm d}(\nu_{\rm rest})$ can be found by extrapolation from other frequencies using $\kappa_{\rm d} \propto \nu^{\beta}$ \citep{chi86}. This emissivity index $\beta$ can be determined from multi-frequency observations covering the entire dust emission spectrum. However, such observations are scarce for high redshift objects, and assume that the dust emission is dominated by a single temperature component. Because both $\beta$ and $T_{\rm d}$ determine the shape of the dust emission spectrum, their values are coupled \citep{pri01}. 
The published values of $\beta$ vary between $\beta=1.3$ for $T_{\rm d}=36$~K \citep{dun00} and $\beta=1.95$ for $T_{\rm d}=41$~K \citep{pri01}. We shall adopt the intermediate value of $\beta=1.5$ for $T_{\rm d}=50$~K \citep{ben99}. These values are consistent with our 113~GHz to 352~GHz measurements (Fig.~\ref{B3SED}).

The value of $\kappa_{\rm d}$ depends on the composition of the dust grains. The estimates of $\kappa_{\rm d}$ normalized to 850~$\mu$m using $\beta=1.5$ vary between $\kappa_{\rm d}(850\mu$m) = 0.04~m$^2$kg$^{-1}$ \citep{dra84} and $\kappa_{\rm d}(850\mu$m) = 0.3~m$^2$kg$^{-1}$ \citep{mat89}. We shall normalize to an intermediate value of $\kappa_{\rm d}(850\mu$m) = 0.11~m$^2$kg$^{-1}$ \citep{hil83}.

Using these values and $S_{850\rm \mu m}=14.15 \pm 1.65$~mJy \citep{reu03c}, we find $$M_{\rm d} \approx 9 \times 10^7h_{65}^{-2} \bigg(\frac{0.11}{\kappa_{\rm d}(850\mu{\rm m})}\bigg) F(T_{\rm d})~{\rm M}_{\odot},$$ with $F(T_{\rm d})=(\exp[h\nu/kT_{\rm d}]-1)/(\exp[h\nu/k\cdot 50{\rm K}]-1)$.

Comparing with the molecular gas mass estimate, this yields a gas-to-dust ratio $$\frac{M(\rm {\rm H_2})}{M_{\rm d}} \approx \frac{850}{F(T_{\rm d})}\bigg(\frac{0.11}{\kappa_{\rm d}(850\mu{\rm m})}\bigg).$$
This is comparable with values found in other HzRGs \citep{pap00} and ULIRGs \citep{san91}.
Alternatively, we can also compare the CO and 850$\mu$m fluxes, yielding a comparison that is independent of the assumed parameters. We find $\Delta V S({\rm CO})/S_{800\mu{\rm m}} = 92$~\kms, again similar to the values found in the 2 HzRGs observed by \cite{pap00}, the $3\sigma$ limit in 53W002 \citep{all00}, and high redshift quasars \citep{cox02b}. This indicates a standard gas/dust ratio in B3~J2330+3927.

\subsection{Ionized gas and associated \HI\ absorption in \Lya}
Figure~\ref{B3Lya2D} shows the complex velocity structure of the \Lya\ line. The emission spatially extends well beyond the radio lobes. This has been seen in several other HzRGs \citep[\eg][]{deb01,vil02}, and excludes shocks as the ionization mechanism at this position.
Because we do not detect a discontinuity in the \Lya\ profile at the position of the radio lobes, this implies that photo-ionization by the central AGN is the dominating ionization mechanism of the \Lya\ line. 

\begin{figure}[ht]
\psfig{file=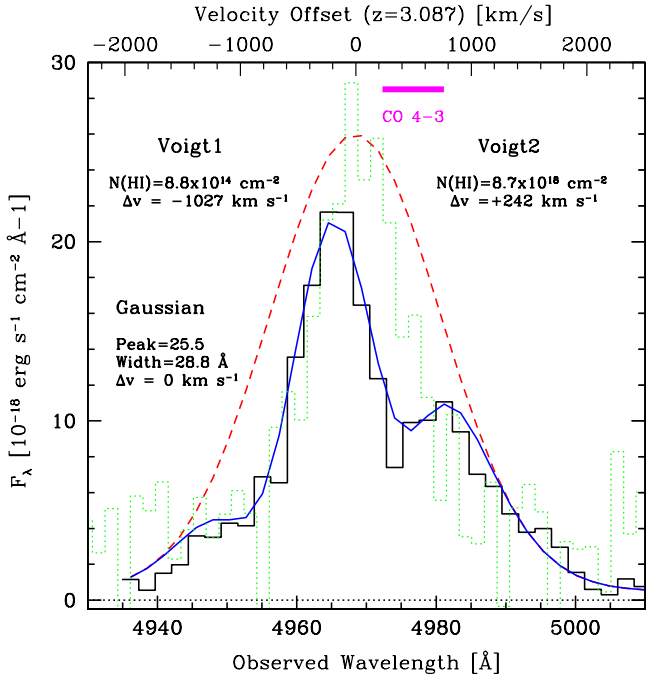,width=9cm}
\caption{\Lya\ velocity profile of B3~J2330+3929 (solid/black histogram). The solid/blue line is the model consisting of a Gaussian emission (dashed/red line) and two Voigt absorption profiles with the indicated parameters. The dotted/green histogram is the scaled \HeII\ profile. The horizontal bar indicates the full velocity width of the CO $4-3$ line (Fig.~\ref{B3COposa}).}
\label{b3Lyafit}
\end{figure}

Figure~\ref{b3Lyafit} zooms in on the \Lya\ line in the Keck/LRIS spectrum (extraction width $1\arcsec \times 1\arcsec$). To check if the velocity profile can be explained by pure velocity structure without absorption, we have attempted to fit the \Lya\ profile with two or three Gaussian components, but this does not provide a good fit between the two peaks, or on the blue wing. 
An alternative interpretation is that the double-peaked structure is caused by associated \HI\ absorption. Such absorbers are often seen in the \Lya\ lines of HzRGs \citep[\eg][]{oji97a,deb00b,jar02}.

We therefore modeled the profile with a Gaussian emission profile, supplemented by two Voigt absorption components (a second component is needed to fit the non-Gaussian blue wing). Figure~\ref{b3Lyafit} presents our best fit, obtained through a $\chi^2$ minimization. We find that the profile can be closely represented with this model. Although the central wavelength of the Gaussian emission was a free parameter in our fit, it corresponds very well with the redshift obtained from the \HeII\ line, providing a strong physical base for our model.

We can estimate the mass of the associated \HI\ absorber using 
\begin{equation}
M(\HI) = m_{\rm p} R^2_{\rm abs} N \quad {\rm kg},
\end{equation}
where $m_{\rm p}$ is the proton mass, $R_{\rm abs}$ is the size of the absorption system, and $N$ is the column density derived from the Voigt profile fitting. In convenient astrophysical units \citep{oji97a}, this becomes: 
\begin{equation}
M(\HI)=10^8 R_{35}^2 N_{19} \quad {\rm M_{\odot}},
\end{equation}
where $R_{35}$ is in units of 35~kpc, and $N_{19}$ is in units of $10^{-19} {\rm cm}^2$.

Because the red absorber has a $\sim$10000 times higher column density ($N=8.7 \times 10^{18}$~cm$^{-2}$) than the blue one, we shall only consider the red absorber in the following. From Figure~\ref{B3Lya2D}, we obtain a lower limit of $\simgt 2\arcsec$ for the full spatial extent of the red absorber. Assuming conservatively a size of $R_{\rm abs} \sim 17 h_{65}^{-1}$~kpc, we derive a total mass estimate of the absorber $$M(\HI) \simgt 2 \times 10^7h_{65}^{-2}{\rm M}_{\odot}.$$ 

We can also use the \Lya\ emission to estimate the mass of ionized hydrogen $M(\HII)$.
Following \citet{mcc90}, we assume pure case B recombination at a temperature of $T=10^4$~K. Using the \Lya\ emission corrected for the associated \HI\ absorption (dashed line in Figure~\ref{b3Lyafit}), we obtain $M(\HII)$ using 
\begin{equation}
M(\HII)=10^9(f_{-5}L_{44}V_{70})^{1/2}{\rm M}_{\odot},
\end{equation}
where $f_{-5}$ is the filling factor in units of 10$^{-5}$, $L_{44}$ is the \Lya\ luminosity in units of $10^{44}$~ergs~s$^{-1}$, and $V_{70}$ is the total volume in units of $10^{70}$~cm$^3$.  
Assuming a filling factor of 10$^{-5}$ \citep{mcc90}, and a volume of $6\arcsec \times 2\arcsec \times 6 \arcsec = 1.4\times 10^{69}h_{65}^{-3}$~cm$^3$ (see Fig.~\ref{B3Lya2D}), we find $$M(\HII) \approx 2.5\times 10^8h_{65}^{-1/2}{\rm M}_{\odot}.$$  This value is consistent with those found in other HzRGs \citep[\eg][]{oji97a}.

\subsection{\HI\ 21~cm absorption against the radio source}

We can determine an upper limit of the \HI\ mass from our non-detection of the \HI\ 21cm line. 
The optical depth is defined as 
\begin{equation}
\tau = -\ln (1- \Delta S/(S c_{\rm f})),
\end{equation}
where $\Delta S$ is the \HI\ absorption flux, $S$ the continuum flux and $c_{\rm f}$ covering factor that is assumed to be 1. Assuming a 3$\sigma$ upper limit, we find $\tau < 1.3$\%.

A limit to the integrated column density can then be derived from the expression \citep[\eg][]{roh86}: 
\begin{equation}
N_{\rm H} \simeq 1.83\times 10^{18} T_{\rm spin} \int \tau \,dv\quad {\rm cm}^{-2},
\end{equation}
where $T_{\rm spin}$ is the spin temperature in Kelvin, and we assume a profile width of 300\kms. \HI\ studies of high-redshift damped \Lya\ systems \citep{car96,kan03} have found values as high as $T_{\rm spin}=1000$~K, but there also appears to be a trend for lower $T_{\rm spin}$ in the most luminous galaxies. We therefore adopt the $T_{\rm spin}=100$~K value found in Galactic clouds \citep[\eg][]{bra92}.
This yields $$N_{\rm H} \simlt 9.3 \times 10^{20} (T_{\rm spin}/100)\quad {\rm cm}^{-2}.$$ Assuming that the absorption occurs only against the radio core, we obtain an estimate of the scale-length of the absorber of $R_{\rm abs}\sim 0\farcs5 \approx 4$~kpc. Using equation (4), this yields $$M(\HI) \simlt 1 \times 10^8h_{65}^{-2}{\rm M}_{\odot}.$$ 

\begin{table}
\caption{Mass estimates of gas and dust components in B3~J2330+3927, subject to uncertainties discussed in \S4.}
\label{massestimates}
\small
\begin{tabular}{rrr}
\hline
Component & Mass estimate & Determined from \\
 & $M_{\odot}$ & \\
\hline
H$_2$ & $7 \times 10^{10}$ & CO line \\
Dynamical & $<6.4\times 10^{10}(\sin^2 i)^{-1}$ & CO line \\
Dust  & $9 \times 10^7$ & 850$\mu$m continuum \\
\HI   & $2 \times 10^7$ & \Lya\ absorption \\
\HII  & $2.5\times 10^8$ & \Lya\ emission \\
\HI   & $<2 \times 10^8$ & \HI\ 21~cm absorption \\
\hline
\end{tabular}
\normalsize
\end{table}

\section{Discussion}

\subsection{The interaction between the different gas components}

We now examine the relationship between the different components summarized in table~\ref{massestimates}.
The similarity of objects~{\em b} and {\em c} in both size and $K-$band magnitude suggests that object~{\em c} may also be part of the same system. We may therefore witness the accretion of the smaller companion object~{\em b} (and possibly also object~{\em c}) by a massive galaxy hosting an AGN (object~{\em a}, but see \S3.2), surrounded by a large reservoir of molecular gas and dust. 
Such gas reservoirs are often directly seen in the huge \Lya\ haloes, which extend out to $>$100~kpc from the AGN \citep[\eg][]{oji96,ven02,vil02,reu03b}. From Figure~\ref{B3Lya2D}, we find that the extent of the \Lya\ halo in B3~J2330+3927 is $\sim$6\arcsec\ (50~kpc), but our spectrum is not very deep, so this value should be considered as a lower limit.

The peaks of the CO, dust and \Lya\ emission are all at object~{\em a}, suggesting that the ionized hydrogen and the CO and dust emission are co-spatial.
Figure~\ref{b3Lyafit} shows that the CO emission also coincides very closely in velocity space with the main associated \HI\ absorber. Because these components are both co-spatial (at object~{\em a}), and have similar velocity shifts, this strongly suggests that we detect CO emission from the same cloud that absorbs the \Lya\ emission.
Our mass estimate from the associated \Lya\ absorber is $\sim 3500$ times lower than the one obtained from the CO emission (Table~\ref{massestimates}). Although both estimates are uncertain to almost an order of magnitude, this difference does indicate that the associated \Lya\ absorber probes only a tiny part of the total gas mass, either because most of the hydrogen is in molecular form, or due to the requirement of a strong background source to detect the \HI\ in absorption.

Our multi-wavelength data of B3~J2330+3926 show for the first time a direct relationship between the associated \HI\ absorption and CO emission in a HzRG. 
High resolution spectroscopy of the \Lya\ and \CIV\ lines of two HzRGs \citep{bin00,jar02} have shown that the absorbers cannot be co-spatial with the emitting gas. The absorption should therefore occur further outward, possibly in a shell pressurized by the expanding radio source \citep{bin00}. Our detection of CO emission from the absorbers suggests that this cloud may be quite extended, but it cannot be primordial, as it is already chemically enriched.
This is also consistent with the recent findings of \citet{bak02}, who found a strong correlation between the equivalent width of \CIV\ absorbers and optical continuum slope in radio-loud quasars. They interpret this relation as evidence for the presence of dust in the associated \CIV\ absorbers, which reddens the optical continuum. The gas and dust emission often trace each other \citep[\eg][]{pap00}, suggesting that the associated absorbers seen in both quasars and radio galaxies probe the large gas and dust reservoirs surrounding AGNs. 

The presence of CO from these reservoirs implies that they contain processed material. This metal emission is sometimes also seen in emission in extended haloes of \CIV, \HeII\ and \NV\ \citep{max02,vil03}. They may have been deposited by previous merger events, or by starburst-driven superwinds at even higher redshifts \citep[\eg][]{tan01a,daw02,aji02,fur03}. Indeed, \citet{tan01b} discuss evidence that a superwind occurs in a $z=3.09$ ``blob'' of \Lya\ emission which also emits strongly at submm wavelengths \citep{cha01}. Such objects may be related to HzRGs, \eg\ when observed during a radio-quiet phase.

To summarize, our observations of B3~J2330+3927 provide further evidence that HzRGs are massive galaxies, located in a dense, interacting environment. Deep near-IR imaging with Keck \citep{wvb98,deb02}, HST \citep{pen01}, and ground-based adaptive optics \citep{stein02} shows that the host galaxies are often surrounded by fainter clumps, like objects~{\em b} and {\em c} in B3~J2330+3926. These clumps may well be concentrations within the large gas/dust reservoir surrounding the AGN, which will eventually merge with the massive central galaxy hosting the AGN.

\subsection{Star formation rate}
Having found evidence that the host galaxy of B3~J2330+3927 is surrounded by a large gas reservoir, which has already been chemically enriched, it is of interest to estimate the ongoing star formation rate in this galaxy. We cannot use the rest-frame UV and optical emission lines, because they are mainly ionized by the AGN, and not by massive stars. Similarly, the UV continuum emission is likely to be contaminated by a scattered quasar component \citep[\eg][]{ver01a}. We therefore determine an upper limit to the global star formation rate of the entire system using the far-IR (FIR) dust emission.

We can determine the total FIR luminosity by integrating the thermal spectrum:
\begin{equation}
L_{\rm FIR}=4 \pi M_{\rm d} \int_0^{\infty} \kappa_{\rm d}(\nu) B(\nu, T_{\rm d}) d\nu,
\end{equation}
yielding: 
\begin{equation}
L_{\rm FIR}=\frac{8\pi h}{c^2}\cdot\frac{\lambda^{\beta}\kappa_{\rm d}(\lambda)}{c^{\beta}}\Bigg(\frac{k T_{\rm d}}{h}\Bigg)^{\beta+4} \Gamma[\beta + 4]\zeta[\beta+4] M_{\rm d},
\end{equation}
where $\Gamma$ and $\zeta$ are the Gamma and Riemann $\zeta$ functions, respectively.

Substituting equation (3) and $x=h\nu_{\rm rest}/k T_{\rm d}$, we find:
\begin{equation}
L_{\rm FIR}=4\pi\Gamma[\beta + 4]\zeta[\beta+4] D_L^2 x^{-(\beta + 4)}(e^x-1)S_{\rm obs}\nu_{\rm obs},
\end{equation}

For the values adopted in this paper, we find: $$L_{\rm FIR} \approx 3.3 \times 10^{13}h_{65}^{-2}{\rm L}_{\odot}.$$

In \S 3.5, we have argued that the dust emission probably consists of a central component near the AGN, and a spatially extended component. This would imply that the power source of the FIR emission is a combination of direct heating by the AGN (heating the central component) and by recently formed massive stars in a starburst (heating the extended component). It is unlikely that the AGN can power the dust emission out to several tens of kpc, while the detection of several companion objects in the optical and near-IR images suggests the presence of stars out to $\sim$40~kpc.

Following \citet{omo01}, we can then calculate the star formation rate 
\begin{equation}
SFR=\delta_{\rm MF}\delta_{\rm SB}(L_{FIR}/10^{10}L_{\odot})~{\rm M}_{\odot}{\rm yr}^{-1},
\end{equation}
where $\delta_{\rm MF}$ is a function of the stellar mass function, and $\delta_{\rm SB}$ is the fraction of the FIR emission heated by the starburst.
Assuming conservatively $\delta_{\rm MF}=1$, we find 
$$SFR \approx 3000 \delta_{\rm SB} h_{65}~{\rm M}_{\odot}{\rm yr}^{-1}.$$

\cite{pap00} report star-formation rates up to 1500~${\rm M}_{\odot}{\rm yr}^{-1}$ in the 2 other HzRGs where CO emission has been detected. Because $\delta_{\rm SB}$ is likely to be significantly smaller than 1, our values are probably lower. Nevertheless, they are still extremely high, indicating that we are witnessing a major starburst phase, possibly triggered by the interaction and/or merging of the different nearby companion objects seen in the $K-$band image.

\section{Conclusions}
The main results from our multi-frequency observations of B3~J2330+3927 are:

$\bullet$
We detect the CO~$J=4-3$ line with a full velocity width of $\sim$500~\kms\ at the position of the AGN host galaxy. The line is centered at $z=3.094$, much closer to the central velocity of the associated \HI\ absorber in \Lya\ than to the systemic redshift determined from the \HeII\ line. This strongly suggests the CO emission and associated \HI\ absorption both originate from the same gas reservoir surrounding the HzRG.

$\bullet$
We find a substantial discrepancy between our 250~GHz single dish and 230~GHz interferometer measurements of the thermal dust emission. This strongly suggests that the dust emission is distributed at spatial scales larger than the  $1\farcs9 \times 1\farcs5$ synthesized interferometer beam. The combined uv-plane data can be fit by a circular Gaussian profile with a size between 0\farcs5 and 5\arcsec.

$\bullet$
The optical/near-IR spectroscopy suggests that the AGN is located at the brightest $K-$band source (object~{\em a}). This is supported by the CO and dust emission, which both peak at the same position. However, the radio morphology suggests that the AGN is located at the central radio component, 1\arcsec\ south of object~{\em a}. If the AGN is located at object~{\em a}, the radio source is extremely asymmetric. Alternatively, the AGN may be located at the central radio component, and could be either heavily obscured in the optical/near-IR, or could be located off-centre from the host galaxy. However, this alternative appears unlikely given the position of the dust/CO peak.

$\bullet$
We obtain a $\tau < 1.3$\% limit on the \HI\ 21~cm absorption against the radio source. This represents the seventh non-detection out of 8 radio galaxies where \HI\ 21~cm absorption has been searched.

$\bullet$
We estimate the mass of the different components in B3~J2330+3927, which indicates that the CO emission traces the largest fraction of the total mass in the system.

In summary, our multi-wavelength observations suggest the presence of a large gas and dust reservoir surrounding the host galaxy of B3~J2330+3927. For the first time, we have found strong indications that these different observations may trace the same material, both in emission and in absorption. The detection of CO emission therefore implies that a substantial part of this reservoir is not primordial, but must have been previously enriched, possibly by a starburst-driven superwind.

\begin{acknowledgements}
We thank Alexandre Beelen and Richard Wilman for their help with MOPSI and the \Lya\ profile fitting, Dennis Downes for useful discussions, and the Plateau de Bure staff for their efficient assistance. We thank the anonymous referee for useful suggestions.
We are indebted to Daniel Stevens for obtaining the initial optical data on this source in 1995.
IRAM is supported by INSU/CNRS (France), MPG (Germany) and IGN (Spain). 
The W.\ M.\ Keck Observatory is operated as a scientific partnership among the California Institute of Technology, the University of California, and the National Aeronautics and Space Administration. The Observatory was made possible by the generous financial support of the W.\ M.\ Keck foundation.
The Westerbork Synthesis Radio Telescope (WSRT) is operated by the Netherlands Foundation for Research in Astronomy (NFRA) with financial support of the Netherlands Organization for Scientific Research (NWO).
The National Radio Astronomy Observatory (NRAO) is operated by Associated Universities, Inc., under a cooperative agreement with the National Science Foundation. 
This work was supported by a Marie Curie Fellowship of the European Community programme 'Improving Human Research Potential and the Socio-Economic Knowledge Base' under contract number HPMF-CT-2000-00721.
The work of DS was carried out at the Jet Propulsion Laboratory, California Institute of Technology, under contract with NASA.
The work by MR and WvB at IGPP/LLNL was performed under the auspices of the U.S. Department of Energy, National Nuclear Security Administration by the University of California, Lawrence Livermore National Laboratory under contract No. W-7405-Eng-48. 
This work was carried out in the context of EARA, the European Association for Research in Astronomy.
\end{acknowledgements}

\end{document}